# Applications for mobile devices focused on support for autism spectrum disorder population and / or people in their immediate environment in their daily lives: a systematic and practical review from a Spanish-speaking perspective

Francisco Crespo • Estefanía Martin
Department of Ciencias de la Computación, Arquitectura de Computadores, Lenguajes y Sistemas Informáticos, Estadística e Investigación Operativa, Universidad Rey Juan Carlos,
Tulipán St., Móstoles City, 28933, Spain
e-mail: hf.crespo@alumnos.urjc.es; estefania.martin@urjc.es
Phone: +34678332030

Abstract There are countless applications for mobile devices of common use (smartphones, tablets, laptops, smartwatches), within which the present study has found 135 applications with an Spanish version available in the digital market (in a practical sense) focused on autism spectrum disorder, developed mostly for the support in daily life of people with autism and / or people from their immediate environment. The main objective of this article is the review of scientific publications on this type of applications. Most of the applications studied focus on communication, social behavior and learning, which coincides with what is observed in the digital market, although very little of the offer has scientific validation. The research and development of these applications from a Spanish-speaking approach is limited. In general, the studies show positive results in terms of learning and permanent adoption of behaviors and skills by using these new technologies. It is recommended to deepen research and further development of applications focused on leisure, resources for parents and professionals, and supporting the needs of adults with autism.

Keywords Autism Spectrum Disorder • Applications • Mobile Devices • Information and Communication Technologies • Spanish

1 Introduction

Autism Spectrum Disorder (ASD) encompasses a specific group of neurodevelopmental disorders, including autism, Asperger's syndrome, childhood disintegrative disorder, and non-specified pervasive developmental disorder. It is defined by that is known as a "triad of alterations" that affect social communication, social interaction and manifest with repetitive or restricted behaviors, in different degrees: mild, moderate and severe (DSM-5 2013). Other currents consider that the triad is composed, in addition to deficits in communication and social interaction, by a deficit in the social imagination (Wing 1992). The American Psychiatric Association suggests using the general label of Autism Spectrum Disorder-ASD, to the detriment of mentioning the specific disorder of neurodevelopment of an individual (Kenny et al. 2015).

According to data from the Center for Disease Control and Prevention of Atlanta (CDC), 1 in 68 children have been diagnosed with ASD, a prevalence that is statistically increasing. ASD is 4.5 times more common in men than in women (CDC 2017). In Latin America there are studies on the prevalence of ASD in countries such as Argentina, Venezuela and Mexico that yield data between 13 to 19/10000 (Elsabbagh et al. 2012); in Spain more than 450,000 people have this type of disorder, although not all are correctly identified and diagnosed (Ministerio de Salud, Servicios Sociales e Igualdad 2012). Its causes have not yet been accurately determined, but it is considered to arise largely as a disorder by combined genetic and environmental factors (Amaral 2017). What is clear is that the brain of people with ASD processes information differently from a neurotypical one, also considering that in some cases intellectual disability can occur simultaneously (McKenzie et al. 2016), reason why it is also sought its explanation in some psychological theories and neurological alterations, such as the theory of mind, the dysfunction of sensory integration, concrete thinking and executive functions (Johnson and Myers 2007).

Given such conditions, people with ASD, especially in their first years of life, present difficulties to identify, interpret and produce social behaviors, which are the basis for communication and social cognition, which makes it difficult to determine the intentions, thoughts and emotions from other people (Carpendale and Lewis 2004). Social cognition includes the ability to adequately interpret nonverbal social and emotional cues, such as speech, facial expressions, body movements, among others. In this way, social situations such as cooperative play and empathic attitudes to the reactions of others are minimal or nonexistent, and in the future, they affect notably in adult life (Boucenna et al. 2013). Thus, early educational intervention has a very important role, through behavioral and developmental methods aimed at promoting skills that are considered fundamental, such as joint attention and imitation, communication, symbolic play, cognitive abilities, sharing emotions and regulation (Murray 1997; Ospina et al. 2008). Additionally, it is important to consider guidelines for treatments, such as:

start as soon as possible; daily intervention; family involvement; periodic evaluations and review of objectives; promotion of spontaneous communication; promote skills through cooperative play; acquisition of new competences and application in natural contexts; and prioritize positive behaviors over challenging activities (Narzisi et al. 2014).

Within the various methods and tools of early intervention in people with ASD, information and communication technologies (ICT) have proven to be beneficial, and there have been great advances in research on ICT for education of people with special needs (Konstantinidis et al. 2009). Education in the early stages of ASD has proven useful in coping with difficulties in understanding what other people communicate (Howlin and Asgharian 1999). Thus, there have been innovations in approaches and methods based on ICT for therapies and education of children with ASD, which has recently been included as a focus in ICT focused on social signal processing, whose objective is to provide computers with the ability to detect and understand social signals and communication (Chaby et al. 2012). It has also been included in ICT focused on affective communication, which seeks to model, recognize, process and simulate human affections (Chetouani et al. 2009; Esposito 2009).

In general, ICT are adequate for the visual thinking of people with ASD; can help those who are not verbal; they constitute a convenient auditory material, provide multiple inputs of data suitable for different types of skills; they are adaptable to sensitivity (auditory, tactile, visual) and offer various communication channels (Aresti and García 2014; Fletcher-Watson 2015). Software intended for people with ASD, especially if they are infants, must respect the design for everyone, and of course adapt to their characteristics, according to their needs and abilities, pace of learning and interests. Additionally, its interface must be friendly and motivating, multiformat, with a progress evaluation system, easily configurable. All the above will make the experience a positive process (Lozano et al. 2011).

Within the ICT, given the accelerated growth, accessibility and commercialization of personal mobile devices (smartphones, tablets, ultrabooks), the applications for these devices focused on supporting people with ASD proliferate, mainly in terms of the early development of imitation and joint attention (Piaget 2007), with diverse treatment objectives, highlighting: interactive environments, virtual environments, avatars, serious games and telerehabilitation. The uses of these applications for ASD treatment can be classified according to their main objective: (1) support technologies that counteract the impact of sensory and cognitive alterations of life related to autism; (2) cognitive rehabilitation / remediation seeking to modify and improve the basic deficit in social cognition; (3) special education programs to counteract the difficulties of children with ASD in the acquisition of social and academic skills; (4) support tools and processes for parents, guardians, caregivers and / or professionals. All this, in addition, implies having an adequate induction by the users towards the available ICT to determine the right hardware and software according to each specific need (Boucenna et al. 2013; Fletcher-Watson 2015). In web portals it is possible to find up many applications for mobile devices, covering each of the approaches to support people with ASD (Autismspeaks 2017; Appyautism 2017).

This notorious and important impact of ICT as a support to people with ASD, and specifically applications for mobile devices, constitute the basis of the present investigation that is part of a systematic review, regarding indexed articles referring to applications for mobile devices focused on the permanent support of people with autism in their daily lives. Additionally, a component is added to the research that seeks to relate the review with the reality of the Spanish-speaking community regarding the accessibility (both in quantity and quality) to applications with the mentioned characteristics and a Spanish version that are available in the market.

2 Method

A systematic search was conducted to identify empirical studies that evaluated applications for mobile devices focused on permanent support to people with ASD in their daily lives.

2.1 Search Strategy

Web of the Science, Scopus, Science Direct, PubMed, Medline and PsyInfo Journals online were searched using a combination of the following free-text terms with Boolean operators and truncation: *autism or ASD, APPS or applications, smartphones, mobile, tablet, technology, technological, iPad, capture, interface, communication, computer, web-based, self-monitoring, language builder, software, portable, computer-based, and human-computer,* excluding *virtual reality, fragile X, robot or robotic, LF-ASD and FMRP.* The search was limited to English speaking Journals. The search occurred during the months of October and November 2017.

2.2 Inclusion Criteria

To be included in the review, the study must report data that evaluated the use, development or review of applications for usual mobile devices focused on support to people with autism spectrum disorder in their daily lives; that is, applications that imply their use in everyday activities of a non-formal nature, be they: recurrent situations or skills learning. So, studies of applications for professional therapeutic treatment, detection and / or evaluation of the disorder; those which require additional equipment to the main mobile device (virtual reality, wearable); and those for formal education in the classroom were excluded. In addition, those focused on cases such as high-functioning autism (HFA) and Asperger syndrome (AS) were also excluded since they are not the most common types of autism. Finally, to strengthen the relevance of the studies, they must have been cited at least 10 times.

2.3 Data Extraction

First, a brief bibliometric analysis was conducted to have an overview of the documents with respect to ICT, applications for mobile devices and the ASD, which allows comparative analysis during the discussion with the documents selected to be included in the systematic review. The bibliometric analysis was developed based on four indicators that are detailed in Table 1. The systematic review included studies that were summarized based on (a) number of participants: age and diagnosis; (b) target skills or target knowledge; (c) activities performed; (d) research design; (e) main results or conclusions; and (f) number of citations. Data extraction was performed by the first author and checked by and independent rater for accuracy. No disagreements on extracted data occurred.

Table 1 Bibliometric indicators about the scientific production in the investigation of digital applications and mobile devices to support people with ASD.

| Bibliometric indicators | Description |
| --- | --- |
| Scientific production per year | Evolution of the theme through articles published until 2017. |
| Research area | Most productive research areas. |
| Most productive sources within the theme | Title of the sources and their format. |
| Most productive countries | Countries with the highest production of scientific works. |

3 Results

3.1 Bibliometric analysis

A total of 199 documents were obtained through the search, which have been included in the bibliometric analysis with the indicators detailed in Table 1.

3.1.1 Scientific production per year

Figure 1 shows the scientific production per year about the research of digital applications and mobile devices focused on ASD. There is a progressive increase in production, highlighting a considerable jump from 2015, reaching a peak in 2016 where 48 articles are published (24% of the total).

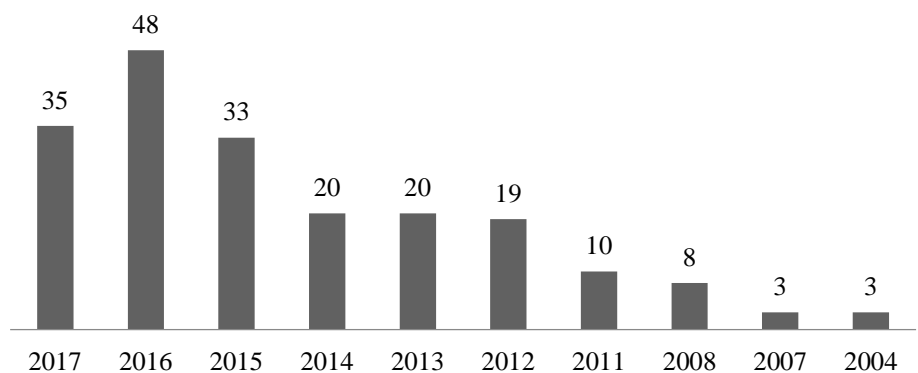

Fig. 1 Scientific production per year

3.1.2 Research Areas

Figure 2 indicates the areas where publications have been made regarding research of mobile applications and devices focused on autism. Health sciences predominate (117), addressing topics such as: general medicine, biology, psychology, psychiatry, neurology, rehabilitation, genetics, pharmacology, physiology, nursing, and occupational health. 74 publications belong to the area of Information and Communication Technologies, and cover technological, computational, engineering, telecommunication and communication approaches. The third and final area is education with 51 publications.

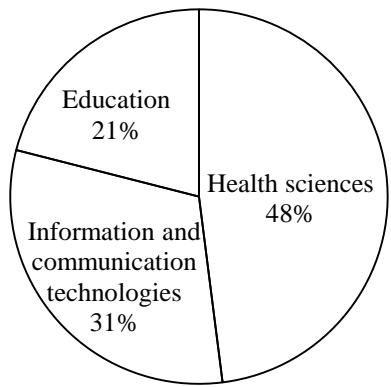
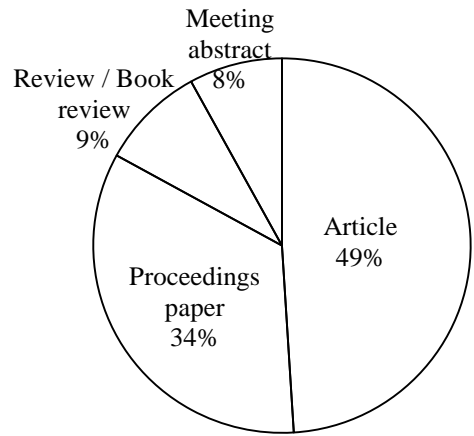

Fig. 2 Research Areas                    Fig. 3 Format Types

3.1.3 Most Productive Sources

Within the 199 publications 4 types of formats were found: articles, proceeding papers, meeting abstracts and reviews. Articles clearly predominate, as can be seen in figure 3. On the other hand, publications come from 249 sources within the four formats, of which 9 of them have at least three publications on the list, which includes 34 articles, 24 proceedings papers and 3 reviews; being the most productive: Lecture Notes in Computer Science (13 articles), Research in Autism Spectrum Disorders (8 articles), Journal of Autism and Developmental Disorders (7 articles) and Journal of Developmental and Physical Disabilities (7 articles); as can be visualized in figure 4.

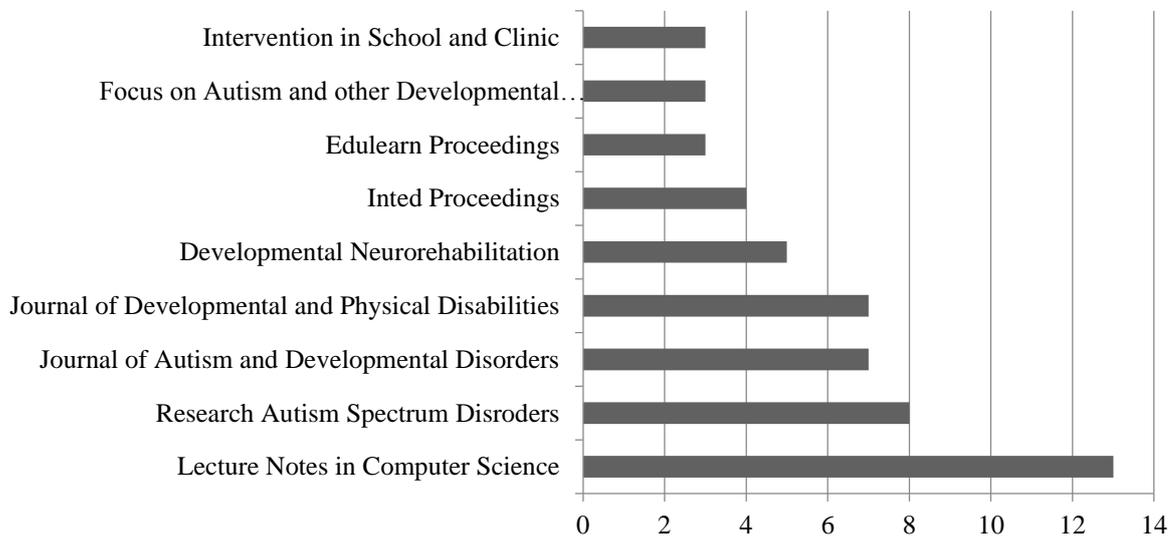

Fig. 4 Most productive sources with 3 or more publications (110 sources with 1 or 2 publications not included)

3.1.4 Most productive countries

Regarding this indicator, the US stands out well above the rest of the countries, with 62 publications that represent 31% of the entire list. England and Malaysia second the list with 11 publications (5.5%) each, followed by Spain with 9 publications (4.5%). France and Italy match in publications with 8 (4%) each. Another group with 7 publications each country is made up of Australia, China and New Zealand. Other countries, 13 in total, cover the remaining 69 publications, with a minimum of 1 publication and a maximum of 6 publications per country. In addition to Spain, the Spanish-speaking countries with publications are Mexico (3) and Ecuador (1).

3.2 Systematic review

For the review, the 199 documents were classified according to the specific focus that they have within the thematic axes that are: digital applications, mobile devices and TEA. In this way, it was found: 74 studies on applications for daily use of a person with ASD and / or their immediate environment (14 of them have been cited more than 10 times); 39 studies focused on formal education (work in the classroom, support for teachers, questionnaires); 30 specific studies (i.e.: based on language or for specific activities such as music teaching or job interviews); 27 studies focused only on mobile devices without considering digital applications; 18 studies focused on professional care (detection, therapies, questionnaires); 17 reviews; and 1 study focused on marketing. Thus, 14 studies met criteria for inclusion in the review. All studies are articles and have been summarized in terms of their aim including those that (a) evaluate apps for acquisition of useful knowledge in daily life; (b) evaluate apps focused on the acquisition of communication skills; and (c) evaluate apps for acquisition of adequate social behavior. These data are presented in Table 2.

3.2.1 Acquisition of Useful knowledge

De Urturi et al. (2011) describe in their study a project that consists in the design and implementation of a multimedia application within the framework of Serious Games. The platform chosen for the development of the application is Android, and it consists of ordering sequences related to first aid, associating images with medical specialties and steps to move inside a hospital or a medical center. Research design consisted in a multiple baseline across participants. The tests were performed on 10 children and adults with ASD, men and women, who have demonstrated capacities to perform activities independently. The results indicate that the application helps enrich and increase educational and therapeutic knowledge, both from a social focus (by increasing motivation with educational games, especially children) and from a medical approach by measuring parameters useful for therapy.

Bereznak et al. (2012) develop a study involving three young men with ASD, between ages of 15 to 18 years. Selection criteria included: IEP goals related to daily living, self-help and vocational skills, generalized motor imitation, ability to attend to a short video, and adequate vision and hearing. A multiple probe across behavior design were used to evaluate the effects of video prompting on skill acquisition (2 individuals used video self-prompting, one used video-prompting). The device used was an iPhone3, and three daily activities were selected: making copies, using a washing machine and making noodles. Participants were given verbal task directions from an instructor. All participants were able to perform a few steps of various behaviors better after the intervention with the iPhone as an effective self-prompting device to teach daily livings and vocational skills. The study concludes that more research is needed before generalizing results across the entire ASD population.

3.2.2 Acquisition of Communication Skills

The first study (De Leo et al. 2011) have designed, developed and tested a software application, called Pix Talk which works on Windows Mobile platform. The tests in this study were conducted under a user-centered design with three children (two boys and a girl) with ASD and their respective teachers through scheduled activity sessions. All participants had experience using PECS. The application consists of the use of images under the PECS system to be able to communicate through a mobile device, the images can be downloaded from a website. The results indicate that in general the app is better than the traditional use of PECS images on paper, only one of the professors showed neutrality when indicating a preference. Regarding limitations, the problem of the battery life of the devices was found and the fact that not all children with ASD are able to use a mobile device properly. The study concludes that the use of computers is beneficial for people with ASD, being more attractive than toys for some children.

Ganz et al. (2013) investigated the efficacy of the PECS Phase III app on an iPad with 3 male children with ASD, complex communication needs (CCN) and prior experience with PECS (two 3-year-old boys and one 4-year-old boy). Research design consisted in a multiple baseline across participants. The study also consisted to investigate the preference between the app and the traditional PECS. Following app instructions, participants independently and correctly used the app thereby establishing a functional relation between app instruction and the post-instruction choice phase, but one of them preferred the traditional PECS communication book.

Sigafoos et al. (2013) worked with two male children with ASD practically non-verbal, aged 4 and 5 years. The objective was to determine the acquisition of requesting repertoire through a multiple baseline across participants research design. Using an iPad and the Proloquo2Go app, activities were carried out to request toys through the app and use them for time intervals (wait-play-wait). The app uses a synthesized voice that vocalizes the physical request (light touch or finger tap). Both children acquired and maintained the ability to order the toys according to the guidelines of the app. One of them decreased his hitting behavior.

Gevarter et al. (2014) evaluated three male pre-school children with ASD and ID with an average of 3 years of age, across three displays in two iPad AAC apps (GoTalk and Scene and Heard). Go Talk includes a Widgit symbol button just like Scene and Heard, but the latter also includes a photographical hotspot. Research design consisted in a multiple baseline across participants. Two of the children more consistently acquired the skills with the photographical hotspot than with the symbol button format but could not master the combined format.

One of the children could similarly master the three conditions. The results of the research would indicate that the applications for portable devices focused on AAC, given their characteristics, can positively influence the requirements skills.

King et al. (2014) worked under a multiple probe design across participants with three children with ASD with limited to no vocal output (2 female, 1 male; between ages of 3 to 5). The study evaluated whether children with ASD can request properly using a speech generating device. An iPad was used to work with Proloquo2Go app. The intervention methodology consisted of an adaptation of PECS system. All participants achieved mastery criteria of PECS Phases 1-3a using the app. The results support the idea that children with ASD, using new technologies, can more quickly and efficiently acquire the skills of requirement of things through AAC, and even improve or acquire the ability of vocal requesting. Technical difficulties of the devices could be a barrier.

Strasberger and Ferreri (2014) worked with four male children with ASD and with no or limited vocal verbal behavior between ages of 5.8 to 12.11 to investigate the acquisition of additional communicative behaviors by using and iPod and Proloquo2Go. A multiple baseline across participants research design was implemented to carry out a PACA training (very similar to PECS). By using the device and the app, all children acquired the ability to request in a complete sentence, three of them could respond to "What do you want?" and two could respond to "What is your name?"

Waddington et al. (2014) developed a study to evaluate a procedure aimed at teaching three male children with ASD and severe communication impairment (7, 8 and 10 years old) to use Proloquo2Go app in an iPad to make general and specific requests for access to toys. Research design consisted in a multiple baseline across participants. The three children improved and maintained their social communication skills using sequences, both with family and non-family members. The study concludes that through systematized instructions, children with ASD can take advantage of applications for mobile devices focused on communication support.

Chien et al. (2015) investigated with the objective of evaluating the iCAN app, a tablet-based system (Android platform) that adopts the aspects of the traditional PECS approach while incorporating digital and portability advantages. The system was deployed under a multiple baseline across participants research design onto eleven children with ASD from moderate to severe and verbal skills from low to none: age range from 5 to 16 years of age (9 male, 2 female). The app reduced content-preparation time by over 70% while also enhancing children with autism willingness to learn and interact with others.

3.2.3 Acquisition of adequate social behavior

Escobedo et al. (2012) present the results of a seven-week deployment study of a mobile assistive app for Android devices, called MOSOCO, that uses augmented reality and the visual support of a validated curriculum to help children with ASD practice social skills in real-life situations. Research design consisted in a multiple baseline across participants. 12 children between ages of 8 to 11 were involved in the study: 3 of them with ASD who demonstrated minimal social skills but age-appropriate functioning and 9 neurotypical (male and female). The app provides with interactive features to encourage making eye contact, maintaining appropriate spatial boundaries, replying to conversation initiators, sharing interests with partners, disengaging appropriately at the end of an interaction, and identifying potential communication partners. Results demonstrate that the app is easy to use and help children practice their social skills. Children with ASD reduced social missteps 56% and behavioral issues 98%; both children increased the number of interaction and their quality 52%.

Hourcade et al. (2012) worked with 26 children over a wide range of the autistic spectrum (5 female, 21 males; ages 5-14). The study conducted a research under a multiple baseline across participants design on computer-based interventions with the goal of promoting social skills. Dell XT2 multitouch tablet was used to work with applications for: drawing with a stylus and fingers, playing music through selecting tiles, and solving visual puzzles. All children enjoyed the social aspects of the apps and demanded more features. The study also suggests that these apps can lead to pro-social behavior, enabling children with ASD to enjoy social activities and express themselves; additionally, apps can help to learn more about children minds and can create safe spaces in which children can explore.

Mintz (2013) presents the results of a qualitative evaluation of an app prototype under a multiple baseline across participants research design, identifying factors that mediate the engagement of fifteen teachers and ten male children with ASD (IQ of 70 or more) with the technology involved. Six parents also participated. The HANDS project aims to improve quality of life for teenagers with an autism diagnosis by providing a mobile ICT toolset supporting them in daily situations (HANDS Project, 2017). This project includes a mobile cognitive support application for smartphones (Windows platform), which supports children with ASD to improve their social and life skill functioning -areas of ability. The study concludes that mobile persuasive interventions for children with ASD are more likely to be effective if the child is both aware of issue and motivated to achieve positive behavior change and remarks that technical issues, like battery charging, must be considering when using high-tech devices.

Murdock et al. (2013) tested the use of the iPad application, called Keynote, to increase the pretend play skills of 4 male children with ASD (4 years old average). Research design consisted in a multiple baseline across participants. With Keynote a series of videos were watched depicting toy figures producing scripted character dialogue, engaged in a pretend play vignette. Three participants increased and maintained the target behavior.

Table 2 Summary of reviewed studies

| Study | Number of Participants (age; diagnosis) | Target Skills / Knowledge | Device/Platform and App | Activities performed | Research design | Main Results or conclusions | Citations |
|---|---|---|---|---|---|---|---|
| De Leo et al, (2011) | 6 (children and adults; 3ASD, 3NT) | Social communication | Windows Mobile / website Pix Talk | PECS training | User-centered design | Computers intervention appear to be particularly appropriate for people with ASD. The app can help to collect communication data in an objective manner. | 14 |
| De Urturi et al, (2011) | 10 (children and adults; ASD) | Handling of situations Healthcare knowledge | Android device 3 generic games | Order sequences Association Moving inside a help center | Multiple baseline across participants | Social and medical benefits. | 23 |
| Bereznak et al, (2012) | 3 (between ages of 15 to 18; ASD) | Vocational and daily living skills | iPhone Video self-prompting | Laundry services Make/prepare a meal or snack Office work | Multiple probe across behavior design | Each participant was able to perform a few steps of various behaviors better after the intervention with the iPhone. | 47 |
| Escobedo et al, (2012) | 12 (between ages of 8 to 11; 3ASD, 9 NT) | Social skills Handling of situations | Android device MOSOCO | Eye Contact Social Boundaries Conversation Share interests Disengage Identify partners | Multiple baseline across participants | Reduced ASD children social missteps 56% Reduce children with ASD behavioral issues 98%. Increase both children the number of interaction and their quality 52%. | 50 |
| Hourcade et al, (2012) | 26 (between ages of 5 to 14; ASD) | Emotions Enjoyable skills | Dell XT2 Generic software (4 apps) | Drawing Music authoring Untangle Photogoo | Multiple baseline across participants | Apps can lead to pro-social behavior, enabling children with ASD to enjoy social activities and express themselves. | 52 |
| Ganz et al, (2013) | 3 (between ages of 3 to 4; ASD and CCN) | Social communication | iPad PECS Phase III | PECS training | Multiple baseline across participants | Following app instruction, participants independently and correctly used the app thereby establishing a functional relation between app instruction and post-instruction choice phase. | 13 |

| Study | Number of Participants (age; diagnosis) | Target Skills / Knowledge | Device/Platform and App | Activities performed | Research design | Main Results or conclusions | Citations |
|---|---|---|---|---|---|---|---|
| Mintz, (2013) | 10 children (15-year-old average; ASD) 15 teachers 6 parents | Social and living skills | Windows Dynamic mobile / HTC HANDS App | Evaluation of results of the app intervention | Multiple baseline across participants | Mobile persuasive interventions for children with ASD are more likely to be effective if the child is both aware of issue and motivated to achieve positive behavior change. | 13 |
| Murdock et al, (2013) | 4 (4 years old; ASD) | Playing skills | iPad Keynote | Playing firefighter | Multiple baseline across participants | There was a clear functional relation demonstrating that the play story intervention had a positive effect on the children. | 24 |
| Sigafoos et al, (2013) | 2 (4 and 5 years old; ASD) | Social communication Emotional Control | iPad Proloquo2Go | Requesting toys and play at intervals of time | Multiple baseline across participants | Correct requests occurring during the final opportunities/interruptions in a percentage of 70 and 75%. | 23 |
| Gevarter et al, (2014) | 3 (3 years old; ASD) | Mand acquisition | iPad Scene and Heard GoTalk | Request items by selecting a symbol or a hotspot on the screen. | Multiple baseline across participants | 2 showed more rapid and consistent acquisition with the photographical hotspot than with the symbol button format. Third participant mastered all 3 conditions at comparable rates. | 11 |
| King et al, (2014) | 3 (3, 4 and 5 years old; ASD) | Requesting skills | iPad Proloquo2Go | Training 4 phases of PECS | Multiple probe design across participants | All participants achieved mastery criteria of PECS Phases 1-3a using the app. | 18 |
| Strasberger and Ferreri, (2014) | 4 (between ages of 5 to 13; ASD) | Social communication | iPod Proloquo2Go | PACA training | Multiple baseline across participants | Research shows that children with an ASD can effectively communicate with an iPod-based SGD. | 12 |

| Study | Number of Participants (age; diagnosis) | Target Skills / Knowledge | Device/Platform and App | Activities performed | Research design | Main Results or conclusions | Citations |
|---|---|---|---|---|---|---|---|
| Waddigton et al, (2014) | 3 (7, 8 and 10 years old; ASD) | Requesting and social communication skills | iPad Proloquo2Go | Access to toys | Multiple baseline across participants | With intervention, all 3 children showed improvement in performing the communication sequence. | 15 |
| Chien et al, (2015) | 22 (11 between ages of 5 to 16, 11 between ages of 26 to 60; 11ASD, 11NT) | Communication skills | Android device iCAN | PECS training | Multiple baseline across participants | The app reduced content-preparation time by over 70% while also enhancing children with autism's willingness to learn and interact with others. | 13 |

One of the participants did not demonstrate progress during intervention and showed general non-compliance. According to the study, a clear functional relation demonstrated that the play story intervention had a positive effect on the children.

4 Applications for mobile devices focused on ASD: current situation in the Spanish-speaking digital market

The research carried out in this regard sought to determine the current situation of the applications focused on ASD for personal devices within the Spanish-speaking digital market to: establish the existing offer and the availability of applications; know the main utilities offered by these applications; and determine the level of scientific validation of them. For the present investigation, it is understood as a Spanish-speaking digital market those portals in Spanish that include in their offer applications with a Spanish version focused on TEA and that also provide adequate tools to be able to select said applications. The above is established considering that only 22% of the Spanish population dominates the English language, according to recent studies (Cambridge University Press 2017; English First 2017).

4.1 Method

A catalog was created with various existing apps in the market oriented to ASD, through a search from a Spanish IP address using (a) the Google search engine that is used by more than 95% of people in Spain (StatCounter 2018), (b) App Store for iPad and iPhone devices and (c) Play Store for Android devices. In this way, 135 applications were selected. The catalog was classified by the following categories: the focus of each application, the platform for which it was developed, the languages, the price, the age of use, and the existence or not of scientific evidence. In addition, a brief description of each app and its website was included.

4.2 Results

4.2.1  Search

Through the App Store, using the word autism in Spanish (autismo), 1963 results were obtained; using the same methodology for the Play Store, 251 results were obtained, in both cases without the possibility of filtering the applications by Spanish version, so they did not comply with the research parameters for the Spanish-speaking digital market. Using the search engine Google, using the phrases "apps para autismo" (apps for autism), "aplicaciones para autismo" (applications for autism), "aplicaciones para trastorno del espectro autista" (applications for autism spectrum disorder) and "apps para trastorno del espectro autista" (apps for autism spectrum disorder), the first two pages of results were reviewed considering habitual behavior the user when doing a search by this means, since the first page of results generates 92% of all the traffic of an average search, and when going from page one to two, the traffic decreases by 95% (Chitika 2013). Both positive and negative results were obtained: applications available in Spanish, applications available without a Spanish version, applications fully developed but not available for download, discontinued applications, and a fully configured portal for a Spanish-speaking person to search and filter applications according to their needs regarding autism named Appyautism, within which 48% of the apps have a version in Spanish, but the predominant language is English (93.5%) although most of the listed applications have versions for more than one language, as it can be seen in Figure 5 that shows the most used languages, taking as a starting point a minimum percentage of 10%. Thus, 135 applications were selected, of which 116 are in the portal Appyautism.

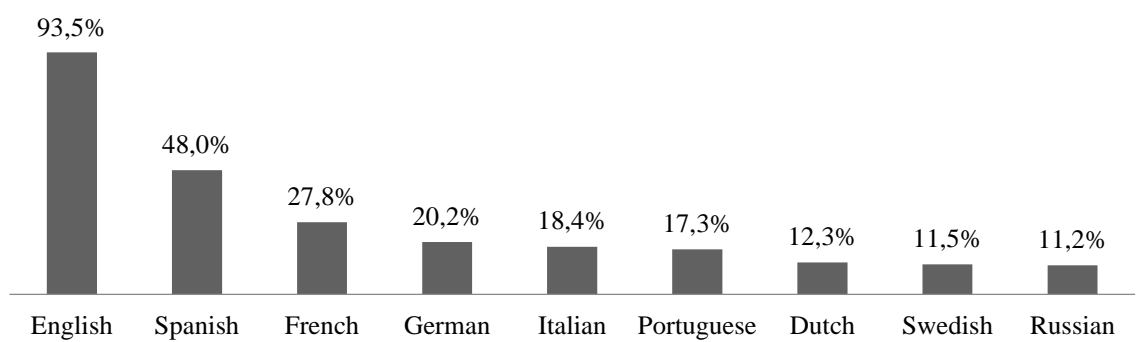

Fig. 5 Languages of list of apps on AppyAutism webpage.

4.2.2  Approach

The applications approach is divided into 6 broad categories: communication, learning, leisure support tools, emotions and social behavior, and resources for parents and professionals. 44% of the applications have more than one focus, however, learning is the most approached (35%) and leisure the least (6%), as can be seen in Figure 6.

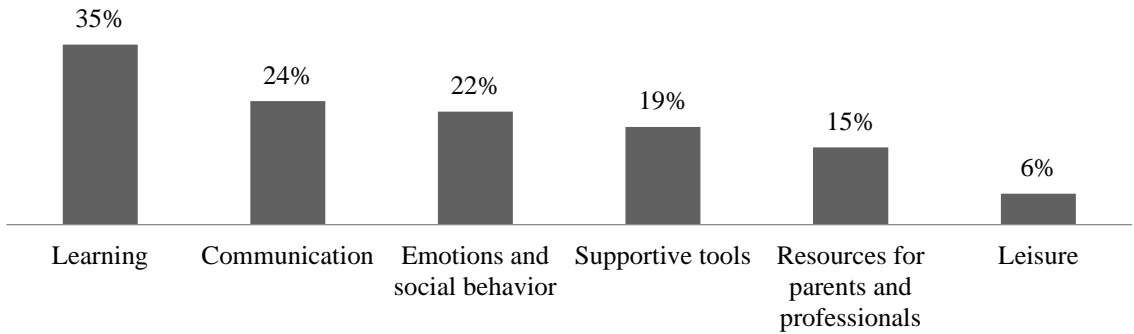

Fig. 6 Approach

### 4.2.3 Platform

Most of applications are developed for iOS-Apple (89.2%), of which 53% are exclusively for this platform. Regarding the Android platform, 42.4% of the applications have a compatible version, of which 9% are developed only for this platform. There are also versions for other platforms such as Windows and web, which can be seen in more detail in Figure 7.

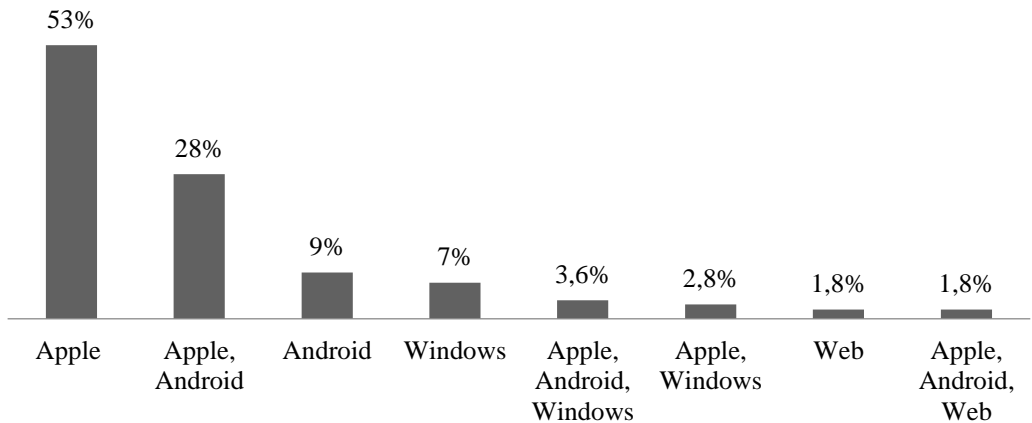

Fig. 7 Platform

### 4.2.4 Price, age of use and scientific validation

Only 32% of the applications are free, the rest varies in price, from 0.89 USD to 1300 USD; many applications are in a price range of 0.89 USD to 10 USD (36%). Regarding the age of use, as can be seen in Figure 8, only 1.4% is aimed at adults and young people (only 0.7% only for adults specifically), 26% is focused on young people and children, and 71% is multipurpose. Only 4% of the applications examined have scientific evidence to support them.

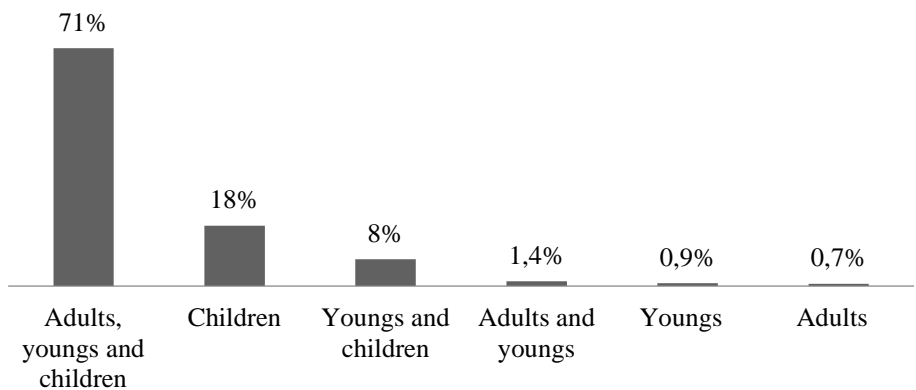

Fig. 8 Ages

5 Discussion

There are several systematic reviews related to applications focused on ASD, this study has made a more detailed approach to those applications whose purpose is to help in daily life of people with autism, be day-to-day activities or common, useful and / or recurrent situations. Similar studies, such as that of Digennaro Reed et al. (2011) focused their review, among other things, on examine the topographies of social skills targeted for intervention in the studies and summarize findings of other relevant variables. Within the selected studies, almost 33% of them focus on the improvement of more than one social skill, within which stands out that 51.7% refers to start a conversation, 27.6% to address skills to play, 20.7% to have a conversation. It remarks there is no study focused on friendship or peer relationships; same thing happens in the present study. The interesting thing about this review is that no study includes as technology: smartphones, iPad/iPods, virtual gadgets or robots; considering that the review was conducted in June 2010, reason why could it be related to a study carried out in 2007 in the United States, which determined a prevalence of 11 cases of autism for every 1000 children (Kogan et al. 2009), which awoke the interest in the subject. Another related systematic review was carried out by Lorah et al. (2015) regarding 17 studies that include the following applications: Proloquo2Go, Pick a Word, My Choice, and Go Talk now; all of them are focused on daily use and were run through an iPad or iPod device; however, this study differs from the present one in that it focuses on the use of portable devices, but not on the applications.

Within the sources researched, the scientific production on digital applications and mobile devices focused on autism has been growing clearly in recent years, being continuous as of 2007. It is curious that, within the studies found, there is almost no production until 2007; the reason for this may also have to do with the previously mentioned prevalence study conducted in 2007 in the United States. Of the 199 studies, only 27 focus solely on mobile devices, 18 on professional attention, 17 reviews and 1 on marketing, so that 136 studies refer to applications with potential to be in the digital market; however, it is a smaller number compared to the applications found on the internet; in fact, within the 14 studies subjected to systematic review 18 applications are analyzed, of which only Go Talk, Scene and Heard, PECS Phase III and Proloquo2Go are within the applications found on internet and only the last one has a Spanish version, forming part of that scarce 4% that have scientific evidence in said list, which could be due to the fact that the development of many of these applications occurs through life experiences, personal or even commercial interests. Demo (2017) presents a critical essay regarding the development of applications focused on autism and its possible deviation towards commercial and even political contexts, as it is an issue that generates more and more public attention. Commercially speaking, almost two thirds of the applications available in the digital market are paid. Regarding life experiences and personal interests, an example is the set of Niki applications, created by a father of a girl with autism (Niki Talk 2017); and although it is true that the development of applications by amateurs or people who have acquired knowledge through their own experiences can generate valid products, there is also the possibility of obtaining useless products (Hamidi et al. 2014) that are placed in the digital market anyway. Thus, it could be said that with a 4% scientific validation, the digital market of applications focused on autism runs the risk of generating distrust in users, since most of the offer lacks a proven scientific research base.

In terms of research areas, within the health sciences there is the highest scientific output in the subject of this study, although not with a marked difference from the areas of ICT and education, with most of publications being articles and proceeding papers (collectively 83%). This is somehow also reflected in (a) the 9 most productive sources, of which 6 belong to the area of health sciences, 2 belong to the ICT area and 1 to the education area; and (b) the 14 articles submitted to systematic review, of which 9 belong to sources within the area of health sciences and 5 to the ICT area.

About the countries with the highest production, the concentration of research in the United States is clearly visible, with works that differ from other countries in quantity, being striking that only 13 out of 199 publications come from Spanish-speaking countries, for which it is not surprising that only 1 article included in the systematic review comes from Spain. This could create drawbacks in the adaptability of the applications, for example, in terms of the sociocultural characteristics of the environments external to the USA, but mainly in relation to the intervention methods of ASD, which may influence the approach of applications to support people with this disorder, since the ABA method (Applied Behavior Analysis) is the recommended and widely adopted in the US, which does not happen in Europe, where other methods are used either for cultural reasons or even for public policies (Keenan et al. 2015). Similarly, observing what happens in the digital market, less than half of the applications found on a webpage focused on Spanish speakers have a Spanish version, and a clear majority use English as their base language, even when they offer the option of selecting another one. Finally, another clear indicator of the predominance of the USA in the research and development of applications focused on ASD are the platforms for which they have been developed; from the study of applications available in the digital market, the predominance of Apple's iOS system is established, which is a leader in the US market, this is also seen in the applications included in the systematic review: 8 of them use the iOS platform (iPhone, iPad, iPod), 3 use Android and 3 Windows Mobile. This represents a limitation in applications aimed at Spanish-speaking people with ASD since, in both Spain and Latin America, the predominant operating system for mobile devices is Android (Moon Technolabs 2017).

Of the 14 studies evaluated in the systematic review regarding applications for mobile devices focused on support for people with ASD and / or people in their immediate environment in their daily lives, 4 of them have used the Proloquo2Go application and 3 have used generic ones; the other applications used are Pix Talk, MOSOCO, HANDS, Keynote, PECS Phase III, Scene and Heard, Go Talk and iCAN. All the reviewed studies focus directly on the person suffering from the disorder and address three general topics: acquisition of useful knowledge, acquisition of adequate social behavior and communication skills, for which various tasks are proposed, ranging from making noodles, going through specific activities of alternative augmentative communication such as the use of PECS system or playful activities such as playing to be a firefighter, to more formal situations such as knowing how to move inside a hospital. 8 of the 14 studies refer to applications that seek the acquisition of communication skills, 4 to applications for the development of adequate social behavior, and 2 to applications for the acquisition of useful knowledge, which can be seen in the digital market, where the most demanded approaches are learning, communication, and emotions and social behavior, in that order. It is striking that 5 studies are based on recreational activities, either directly (De Urturi et al, 2011) or as support for the achievement of the marked objectives (Hourcade et al 2012; Murdock et al. 2013; Sigafoos et al. 2013; Waddington et al. 2014), but their focus is not leisure; in fact, this is also reflected in the digital market of applications to support ASD population, where leisure approach barely represents 6% of it. Only one study includes adults as direct participants of the research (De Urturi et al, 2011); other studies just include them as support elements for the development of the procedures (teachers and parents) (De Leo et al. 2011; Escobedo et al. 2012; Mintz 2013; Waddington et al. 2014), focusing almost entirely on children with ASD with serious communication problems, although two of them (Escobedo et al. 2012; Mintz 2013) condition that participants have characteristics of moderate autism. Into this target group consisting of 97 participants, only 21 (21.6%) are women, which could be related to the fact that ASD is 4.5 times more common in men than in women.

In general, the 14 reviewed studies show positive results regarding the use of applications on mobile devices as support for people with ASD in their daily lives, on the one hand, accelerating and consolidating learning, and on the other generating in users a greater desire to learn and interact with others, with the field of alternative augmentative communication the one that seems to obtain the greatest benefits with the use of new technologies; only one teacher said he could not decide between the application or the traditional use of PECS (De Leo et al. 2011) and one child preferred the traditional PECS communication book (Ganz et al. 2013). Sigafoos et al. (2013) observed the case of a child who, in addition to improving his communication skills, improved his social behavior by decreasing his tendency to hit. Thus, in terms of advantages, aspects such as: dispensing with paper in PECS-type systems, the use of more eye-catching and interactive devices and programs (low-tech vs hi-tech), the ease and convenience of use, the ability to systematization, the possibility of creating safe learning environments and the wide range of resources offered by both applications and mobile devices. MOSOCO application, from the point of view of the everydayness, is the one that more variety of supports presents for the user (Escobedo et al. 2012). On the other hand, three studies point out as a technical limitation the battery life of devices (De Leo et al. 2011; King et al. 2014; Mintz 2013), which could affect users with little expertise or problems to manage frustration. Escobedo et al. (2012) indicate that a limitation of their study was to work on social situations focused on real life in controlled scenarios; in this sense, all the studies reviewed would have this limitation. Murdock et al. (2013) had difficulties with a participant who did not show any progress, but they considered it as a particular case since that participant showed a permanent general disinterest. Practically in all the conclusions it is indicated that more research is necessary to be able to generalize the obtained results.

5.1 Future Directions

The present investigation is focused on studies of applications that are beneficial for people with the most common types of autism, that is why HFA and AS were excluded, considering that HFS constitutes only a fraction (11 to 34%) of the cases of autism, and AS presents many different characteristics (Gillberg 1998). Similar studies are necessary for this type of disorders to determine the applications for mobile devices focused on them. This investigation also considered the immediate environment of the person with ASD: parents, relatives, guardians, caregivers; however, no analyzed study took this group into account as a direct beneficiary of the development of applications, this being another field that must be addressed. Escobedo et al. (2012) remark that a limitation of their work is the fact that it was done in a controlled classroom, although its use is focused on real-life social situations; in fact, it is a limitation in all the analyzed studies, so alternatives should be looked to work in real spaces for everyone. New researches should also include specific studies on the use of applications for mobile devices by women with ASD from the perspective of their gender characteristics, since the condition that prevalence of ASD in men is much higher seems to predispose that studies focus sharply on them.

Taking into account that there are three well-defined poles in terms of the research areas to which publications of the present study belong, an analysis of how related these areas are to each other at the time of developing the applications is relevant, and determining the effectiveness and relevance of these according to their areas because gaps can be generated among the neurological-psychological-behavioral, technical-engineering and educational approaches, if there is not adequate coordination between them.

Regarding approaches to applications in the digital market, almost 50% of these seek to cover several ones that contribute to the daily life of people with ASD; however, this could tend towards a generalization that does not take into account that each person with ASD has her/his own particular characteristics; studies in this regard could propose solutions such as the development of very specific applications or, on the contrary, fully customizable. From the commercial point of view and thinking about the benefit of the user, the scientific community should consider studies of effectiveness of the applications focused on autism available in the market, especially those with greater demand, to be able to overcome that great gap of only 4% scientific validation. On the other hand, the results of the systematic and practical reviews suggest more research and development of applications (a) for adults with ASD, (b) parents and professional and (c) with focus on leisure, since they are groups and a focus practically forgotten in that sense. Of the studies reviewed only one focus on leisure (Murdock et al. 2013).

De Urturi et al. (2011) develop the only research within the review of applications within the framework of Serious Games, which is a field that should be much more exploited and researched to even take advantage of consoles for games and gaming computers, which are increasingly adaptable to mobile devices. Also, in future studies should be considered emerging technologies that little by little will become daily, such as virtual and augmented reality devices, and wearables. Another aspect that should not be overlooked in future studies regarding the use of mobile devices, especially those that involve more body contact, are the sensory difficulties that some people with ASD present, be they tactile, visual or auditory.

Regarding the Spanish-speaking environment, it is necessary to determine the generalities of the applications focused on autism that are developed in the United States, because as mentioned, given the cultural differences and the methods of intervention of ASD, the ways of approaching the development of such applications for mobile devices can vary considerably between the United States and Spanish-speaking countries. On the other hand, a study should not be ruled out regarding the implications when developing specific applications for people with autism in Spanish-speaking countries, considering the particularities of the Spanish language used in each of these countries (Palacios 2006), every time that in the world of autism small details can mean great alterations. But, first, specific studies on applications and mobile devices to support Spanish-speaking people with ASD are essential, as there are already studies on the subject for Chinese, Arabs, Turks, among others.

In short, the present study suggests that the use of apps and mobile devices should be seriously considered when seeking to acquire communication and social skills, as well as social behavior improvements in people with ASD. In addition, it is necessary to carry out more studies to generalize the results and expand the research mainly considering: the whole spectrum of autism, support for adults with ASD, little-seen needs such as leisure, sociocultural aspects, and the use of emerging technologies such as virtual reality devices and wearables.